# Large magneto-thermal-switching ratio in superconducting Pb wires

M. Yoshida,[1] H. Arima,[1] A. Yamashita,[1] K. Uchida,[2] and Y. Mizuguchi[1,a]

[1]*Department of Physics, Tokyo Metropolitan University, 1-1, Minami-osawa, Hachioji, 192-0397, Japan*

[2]*National Institute for Materials Science, 1-2-1, Sengen, Tsukuba, 305-0047, Japan*

**Abstract:** Thermal switching by magnetic fields is one of the important functionalities in thermal management technologies. In low-temperature devices, superconducting states can be used as a magneto-thermal-switching (MTS) component, because carrier thermal conductivity ($\kappa$) is strongly suppressed in superconducting states. Recently, we demonstrated that the MTS ratio (MTSR) of pure Nb reached 650% at a temperature ($T$) of 2.5 K under a magnetic field ($H$) of 4.0 kOe [M. Yoshida et al., Appl. Phys. Express 16, 033002 (2023)]. In this study, to enrich knowledge on MTS of superconductors, the MTSRs of pure Pb wires with 5N and 3N purities were investigated by measuring the temperature or magnetic field dependences of $\kappa$. For 5N-Pb, a large MTSR exceeding 1000% was observed below 3.6 K under $H$ > 600 Oe. Although higher MTSRs were expected at lower temperatures under $H$ > 600 Oe, the obtained data under those conditions were accompanied by large errors due to magnetic-field-induced huge $\kappa$ at low temperatures. In contrast, the $\kappa$ for 3N-Pb were observed to be clearly lower than that for 5N-Pb. Although the magnetic-field-induced change in $\kappa$ was small, the MTSR at $T$ = 2.5 K was 300%. These results suggest that Pb is a promising material for low-temperature magneto-thermal switching because of wide-range $\kappa$ tunable by magnetic field and the purity.

---

[a] Author to whom correspondence should be addressed. Electronic mail: mizugu@tmu.ac.jp.

Superconductivity is a quantum phenomenon characterized by the disappearance of electrical resistivity and the exclusion of magnetic flux at temperatures below a superconducting transition temperature ($T_c$). Zero-resistance states have been used in superconducting high-field magnets and superconducting power cables. Furthermore, Josephson junctions comprising superconductors separated by a non-superconducting thin layer have been used in numerous sensors and quantum computers.[1–5] In addition, thermal management systems utilize superconducting states; for example, high-$T_c$ superconductors have been used in current cables of superconducting magnets owing to its low thermal conductivity ($\kappa$).[6,7] The low $\kappa$ is attributed to the fact that transitions from a normal to superconducting state involve the suppression of carrier thermal conduction. Therefore, high-$T_c$ superconducting bars have zero electrical resistivity and work as thermal barriers in their superconducting states. Magneto-thermal switching (MTS) is another application of superconducting states, where $\kappa$ can be switched by applying magnetic fields that suppress the superconducting states. Although the uses of MTS properties of superconductors are currently limited, they are expected to find applications in low-temperature electronics. Owing to the recent development of high-$T_c$ superconductors using hydrogen-rich compositions, the working temperature is expected to be extended in the future.[8–12] Because MTS[13,14] is one of the important technologies employed in the field of thermal management,[15–17] enriching knowledge on the properties of the MTS of superconductors is important.

In our recent paper,[18] we reported that the MTS ratio (MTSR), calculated as MTSR $(T, H) = [\kappa(T, H) - \kappa(T, H = 0 \text{ Oe})]/\kappa(T, H = 0 \text{ Oe})$, respectively, of pure Nb reached 650% at $T = 2.5$ K under $H = 4.0$ kOe.[18] Herein, $T$ and $H$ denote the temperature and the magnetic field, respectively. Due to the emergence of superconducting states, $\kappa$ decreased with decreasing temperature below $T_c$ of Nb ($T_c = 9.2$ K at $H = 0$ Oe). $T_c$ decreased when magnetic fields were applied, and the carrier contribution to $\kappa$ at a specific temperature was recovered after the suppression of superconductivity. Since Nb is a type-II superconductor, a similar investigation on the MTSR of a type-I superconductor is necessary to enrich knowledge on MTS of superconductors because investigation on MTSR of superconductors has not been reported, except for Ref. 18. Differences in superconducting states should correlate with the MTS properties; hence, studies on various superconductors are desired. Therefore, in this study, we measured $\kappa$ of pure Pb (5N > 99.999% and 3N > 99.9% purities), which is a type-I superconductor with $T_c = 7.2$ K at $H = 0$ Oe, under magnetic fields and observed a large MTSR exceeding 1000% for a 5N-Pb wire below 3.6 K under $H > 600$ Oe. In addition, we evaluated the MTSR of a 3N-Pb wire to investigate the effects of purity on MTS.

The $\kappa$ of the 5N-Pb (polycrystalline, 5N purity, Nilaco) and 3N-Pb (polycrystalline, 3N purity, Nilaco) wires with a diameter of 0.5 mm were measured by a four-terminal configuration using the thermal transport option (TTO) of the physical property measurement system (PPMS, Quantum Design). The terminals were fabricated using Cu wires of diameter 0.2 mm and Ag pastes [see Fig. 1(a)]. To confirm the reproducibility of $\kappa(T)$ for 5N-Pb, two samples were measured; data from sample

#1 are used in the main text, while those from sample #2 are discussed in the supplementary material. The distance between the two thermometer terminals was 23.2, 14.0, and 8.0 mm for the 5N-Pb (sample #1), 5N-Pb (sample #2), and 3N-Pb wires, respectively. We confirmed that the difference in the distance between the two thermometer terminals does not largely affect the $\kappa(T)$ results (Fig. 1 and Fig. S1). Magnetic fields were applied to Pb along the direction perpendicular to the generated temperature gradient [see Fig. 1(a)]. Although we measured $\kappa(T)$ for 5N-Pb (sample #1) at low temperatures ($T < 3.2$ K) under $H > 620$ Oe, reliable data were not obtained due to too high $\kappa$ at low temperatures. Therefore, in this study, we used the data obtained under $H = 620$ Oe as the maximum for 5N-Pb in the temperature-dependence scan. To precisely estimate MTSR at various temperatures, the $\kappa$-$H$ scans were performed at stable field and temperature. A typical measurement period of 30 s below 10 K was considered. Magnetization of the 5N-Pb samples was measured by a superconducting interference device on the magnetic property measurement system (MPMS 3, Quantum Design). Electrical resistivity measurements and Hall measurements were performed by four-terminal methods on PPMS. Carrier concentration was estimated by assuming a single-band model.

Figure 1(b) shows the $T$ dependence of $\kappa$ for 5N-Pb (sample #1) under various magnetic fields up to 620 Oe. Below 7 K, anomalies related to the suppression of carrier contribution to $\kappa$ were observed. With an increase in $H$, the anomaly temperature decreased, and a huge increase in $\kappa$ with a decrease in temperature was observed in the normal states. At low temperatures, $\kappa$ largely increased with decreasing temperature in the normal states. In high-purity metals, a similar increase in $\kappa$ is observed.[19] This behavior (an increase in $\kappa$ by cooling) is generally explained by the suppression of phonon or vibration effects that scattered the electrons, and a peak structure appears as a result of competing influence of electron-phonon and electron-impurity scattering processes. The obtained results were qualitatively consistent with those reported by Quantum Design researchers.[20] Furthermore, the zero-field data were quantitatively consistent with those of a previous study on a Pb crystal (80 mm in length and 6mm in diameter),[21] and the data under magnetic fields are consistent with the previous results ($H = 900$ and 2400 Oe in Ref. 21). To evaluate MTSR, the dependence of $\kappa$ on $H$ at $T = 3.5$ K for 5N-Pb was estimated from $\kappa(T)$ and plotted, as shown in Fig. 1(c). The values of the calculated MTSR ($H$) were also plotted in Fig. 1(d). MTSR rapidly increased under $H > 400$ Oe, and the largest MTSR of 1150% was observed at $T = 3.5$ K under $H = 620$ Oe. As shown in the supplementary material (Fig. S2), no hysteresis of MTS in 5N-Pb was observed (the hysteresis investigation was performed at $T = 3.2$ K). The critical field ($H_c$) phase diagram was designed using the magnetization data presented in the supplementary material (Fig. S3). The huge jump in the $\kappa$-$H$ curve was observed near $H_c$. The $H_c$ value for $T = 0$ K) was estimated to be 803 Oe, suggesting the appearance of a higher MTSR at lower temperatures under fields of $> 803$ Oe.

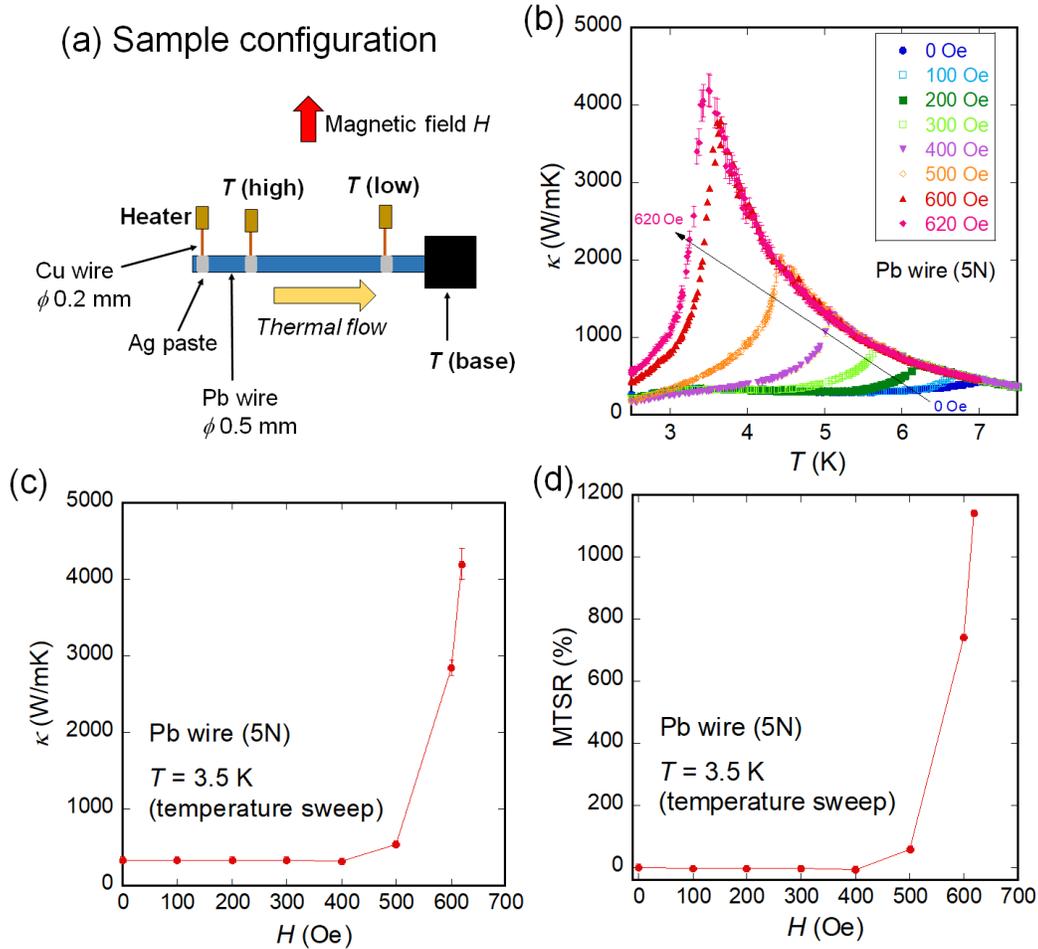

FIG. 1. (a) Schematic image of sample configuration. (b) Temperature ($T$) dependence of $\kappa$ for the 5N-Pb wire (sample #1) under magnetic fields up to 620 Oe. (b) Magnetic-field ($H$) dependence of $\kappa$ at $T$ = 3.5 K; the data were estimated from the $T$ dependence of $\kappa$. (c) $H$ dependence of the magneto-thermal-switching ratio (MTSR) at $T$ = 3.5 K.

To further investigate MTS characteristics of the 5N-Pb wire, $\kappa$ was measured at various stable magnetic fields and temperatures. By optimizing the measurement conditions, we could obtain data up to $\kappa$ = 7500 W/mK. Figures 2(a–c) show the $H$ dependences of $\kappa$ for the 5N-Pb wire at $T$ = 3.2, 3.4, and 3.6 K, respectively. A huge increase in $\kappa$ was observed near $H_c$, and the $\kappa$ values in normal states ($H > H_c$) gradually decreases with increasing $H$. The gradual decrease in $\kappa$ in the normal state is caused by the conventional magnetoresistance due the Lorentz force [see Fig. 1(a) for direction of $H$ and thermal flow]. On the basis of the $\kappa$-$H$ data, MTSR at $T$ = 3.2, 3.4, and 3.6 K was estimated and plotted, as shown in Fig. 2(d–f). At $T$ = 3.2 K, MTSR reaches 2000% under $H$ = 650 Oe, and the highest MTSR systematically decreases with increasing temperature; MTSR ~ 1400% under $H$ = 640 Oe at $T$ = 3.4 K and MTSR ~ 1150% under $H$ = 620 Oe at $T$ = 3.6 K. The huge MTSR will be useful for MTS in devices where a large thermal flow works.

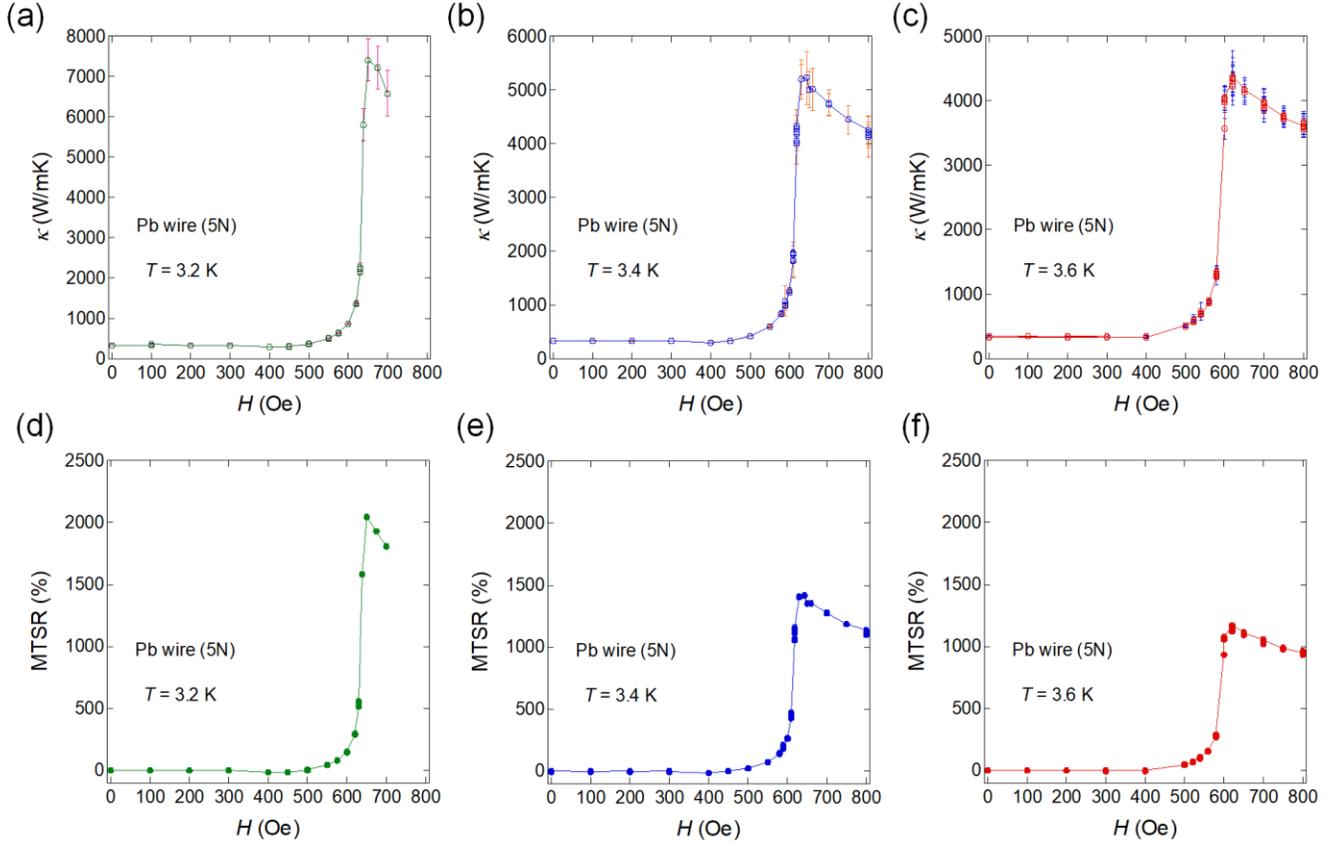

FIG. 2. (a–c) $H$ dependence of $\kappa$ for the 5N-Pb wire at $T$ = 3.2, 3.4, and 3.6 K. (d–f) MTSR for the 5N-Pb wire at $T$ = 3.2, 3.4, and 3.6 K.

Here, we discuss the effects of purity on $\kappa$ and MTS of Pb. In general, purity of metals largely affects thermal resistivity because of the presence of the electron-impurity scattering term, which is proportional to $T^{-1}$, at low temperatures.[19] Therefore, to investigate the importance of purity on the MTS characteristics, $\kappa$-$T$ for the 3N-Pb wire was measured. As shown in Fig. 3(a), decreases in $\kappa$ at $T_c$ was observed, and the $T_c$ values were consistent with those observed for 5N-Pb [Fig. 1(b)]. However, the $\kappa$ values for 3N-Pb were clearly lower than those observed for 5N-Pb. No huge increase in $\kappa$ at low temperatures was observed for 3N-Pb, and the trend is similar to that for Nb.[18] Figure 3(b) shows the $H$ dependence of $\kappa$ for the 3N-Pb wire at $T$ = 2.5 K. Although the magneto-switching value in the 3N-Pb wire is smaller than that in the 5N-Pb wire, MTSR of ~300% was observed at $T$ = 2.5 K. Since the MTSR is high and the $\kappa$ at $H$ = 0 Oe is low for 3N-Pb, low-purity Pb will be suitable for applications where $\kappa$ should be switched between low and moderate values. We observed clear differences in the MTS characteristics and the $\kappa$ values between the 5N-Pb and 3N-Pb wires and ascribed to the difference to the electron-impurity term as mentioned above. To confirm that, low-temperature electrical resistivity ($\rho$), Hall coefficient ($R_H$), and carrier concentration ($n$) were measured and displayed in supplemental materials (Fig. S4). Although there are slight differences in residual resistivity and carrier concentration between 5N-Pb and 3N-Pb, those are not enough to explain the huge difference in $\kappa$. Therefore, the assumed difference in electron-impurity scattering would be essential in Pb.

In conclusion, we measured the values of $\kappa$ for pure Pb wires with 5N and 3N purities under various magnetic fields using PPMS-TTO and estimated their MTSR. For the 5N- Pb wire, MTSR exceeding 1000% was observed below 3.6 K under $H >$ 600 Oe. At higher $H$, we should observe larger MTSR in 5N-Pb samples. Therefore, reliable measurement of larger values of $\kappa$ using other measurement techniques are required to know the maximum MTSR of 5N-Pb or Pb with a higher purity. The MTSR of the 3N-Pb wire was ~300% at $T$ = 2.5 K, which would be caused by short mean free path due to the presence of impurities. Because of different $\kappa$ range in the MTS in Pb, both 5N-Pb and 3N-Pb materials should have each merit for MTS in low-temperature devices.

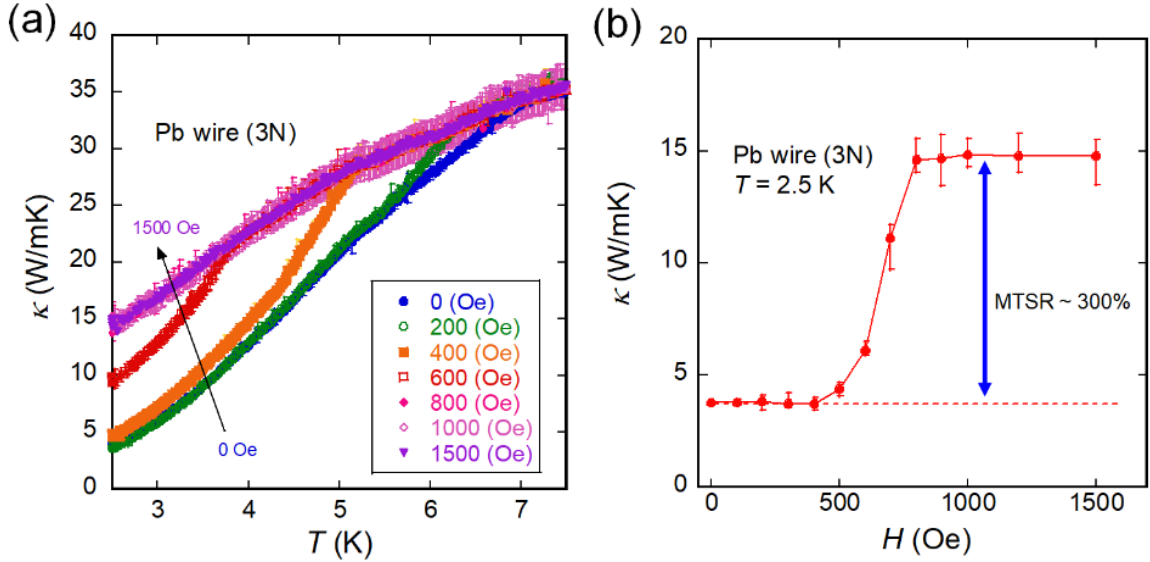

FIG. 3. (a) $T$ dependence of $\kappa$ for 3N-Pb wire under magnetic fields up to 1500 Oe. (b) $H$ dependence of $\kappa$ at $T$ = 2.5 K.

**SUPPLEMENTARY MATERIAL**

See supplementary material for the temperature dependence of thermal conductivity data for 5N-Pb (sample #2), magnetization, resistivity, and Hall coefficient data for Pb.

## AUTHORS' CONTRIBUTION

**M. Yoshida**: Conceptualization, Data curation, Formal Analysis, Investigation, Visualization, Writing – original draft

**H. Arima**: Data curation, Formal Analysis, Investigation, Visualization

**A. Yamashita**: Investigation, Supervision, Writing – review & editing

**K. Uchida**: Conceptualization, Investigation, Project administration, Supervision, Writing – review & editing

**Y. Mizuguchi**: Conceptualization, Data curation, Formal Analysis, Funding acquisition, Investigation, Methodology, Project administration, Resources, Supervision, Validation, Visualization, Writing – original draft

### Conflict of Interest

The authors declare no competing interests.

## ACKNOWLEDGMENTS

The authors thank O. Miura and Md. R. Kasem for supports in magnetization experiments and F. Ando, Y. Oikawa, H. Nagano, and Y. Hirayama for valuable discussions. The work was partly supported by JST-ERATO (JPMJER2201) and Tokyo Government Advanced Research (H31-1).

# Large magneto-thermal-switching ratio in superconducting Pb wires

M. Yoshida,[1] H. Arima,[1] A. Yamashita,[1] K. Uchida,[2] and Y. Mizuguchi[1]

(Corresponding author: mizugu@tmu.ac.jp)

[1]*Department of Physics, Tokyo Metropolitan University, 1-1, Minami-osawa, Hachioji, 192-0397, Japan*

[2]*National Institute for Materials Science, 1-2-1, Sengen, Tsukuba, 305-0047, Japan*

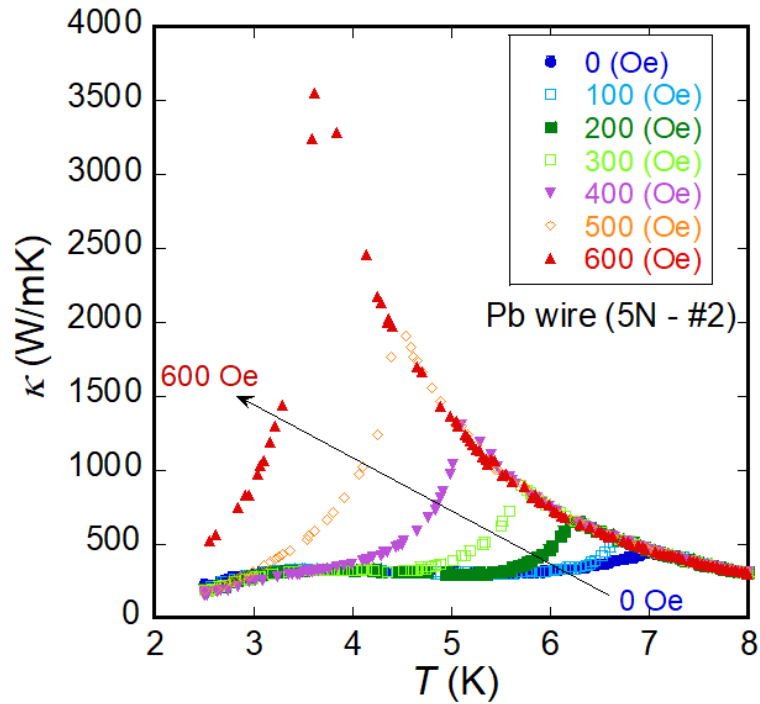

FIG. 1. Temperature ($T$) dependence of the thermal conductivity ($\kappa$) for the 5N-Pb wire (sample #2) under magnetic fields up to 600 Oe.

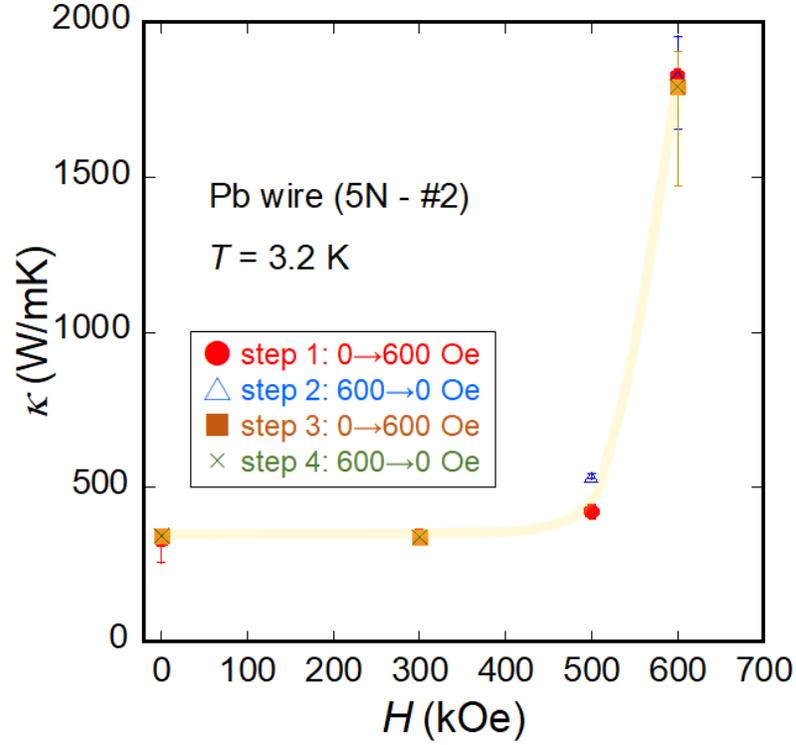

FIG. S2. Magnetic field ($H$) dependence of $\kappa$, measured with two field cycles. Step 1 is the initial increase of field from 0 to 600 Oe, followed by a decrease from 600 to 0 Oe in step 2. Step 3 and 4 are repetition after step 1 and 2. No hysteresis was observed. The yellow line is an eye guide.

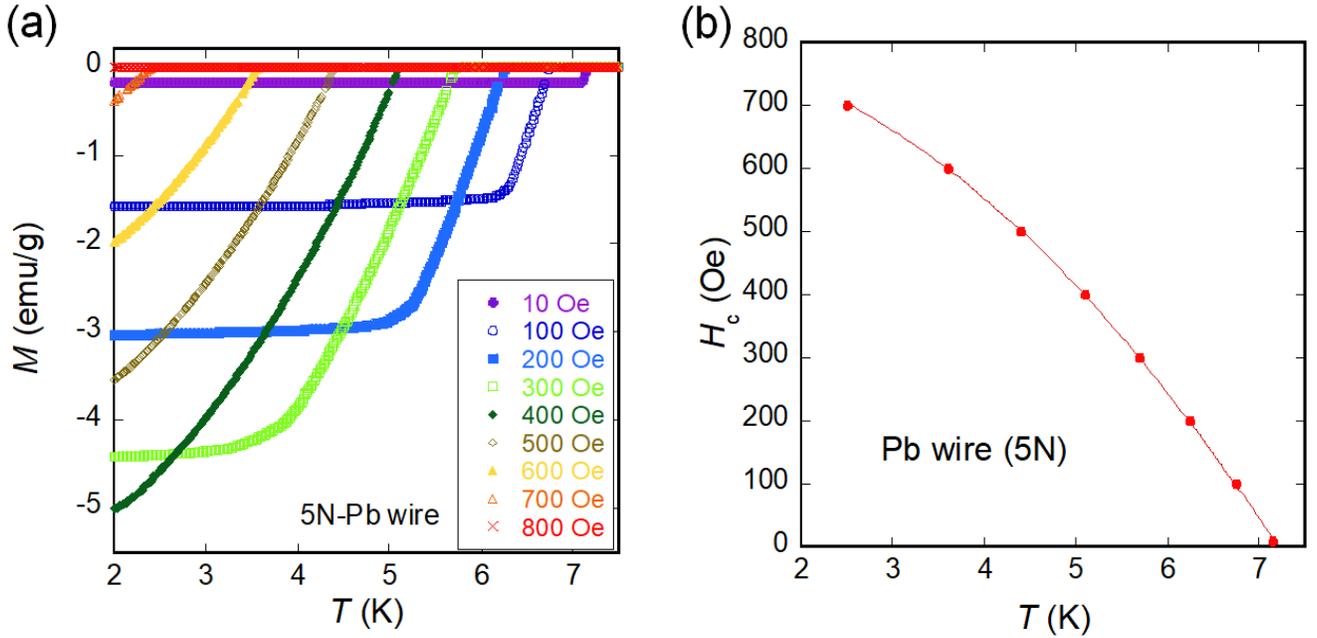

FIG. S3. (a) $T$ dependence of the magnetization ($M$) for the 5N-Pb wire under various values of the magnetic fields. (b) Critical field ($H_c$)-$T$ phase diagram for the 5N-Pb wire.

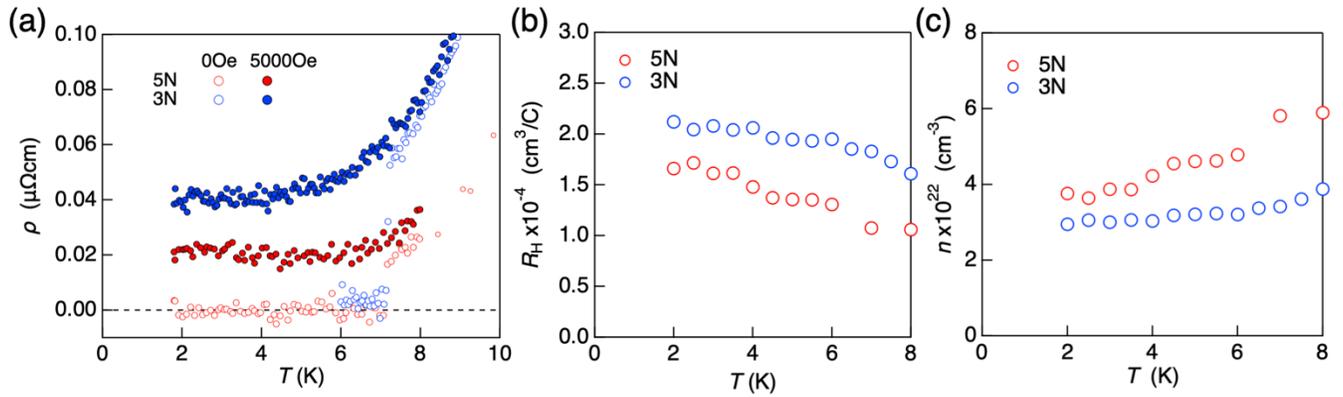

FIG. S4. (a) $T$ dependence of the electrical resistivity ($\rho$) for the 5N-Pb wire and the 3N-Pb sheet under $H = 0$ and 5000 Oe. (b, c) $T$ dependence of the Hall coefficient and estimated carrier concentration for the 5N-Pb and 3N-Pb samples.